\documentclass[11pt]{article}

\usepackage{amsmath}
\usepackage{graphicx}
\usepackage{indentfirst}
\usepackage{amssymb}
\usepackage{cite}
\usepackage{color}
\usepackage{subfigure}
\usepackage{xcolor}
\usepackage[breaklinks=true]{hyperref}

\usepackage{mathrsfs}
\usepackage{amsmath}

\begin{document}

\title{Cosmological Evolution of the Randall-Sundrum II Model with Running Vacuum: A Special Class of Solutions}
\date{}
\maketitle

\begin{center}
	\author{Hao Yu$~^{}$\footnote{yuhaocd@cqu.edu.cn},
		Jin Li~$^{}$\footnote{cqujinli1983@cqu.edu.cn},
Zi-Chao Lin$^{}$ \footnote{linzch24@cqu.edu.cn, Corresponding author}}
\end{center}

\begin{center}
 Physics Department, Chongqing University, Chongqing 401331, China\\
\end{center}

\begin{abstract}
In this work, we construct a novel cosmological framework by integrating the running vacuum model into the Randall-Sundrum II braneworld scenario. We derive a special class of analytical solutions in the model. Our investigation focuses on the complete evolutionary trajectory of the universe, with particular emphasis on the realization of primordial inflation and the evolution of cosmic entropy. We evaluate the model's viability across three scenarios: dust-dominated, radiation-dominated, and general perfect fluid-dominated universes. Our findings show that while the hybrid model is in principle capable of supporting inflation, producing enough e‑fold number demands either severe fine‑tuning of the parameter or unreasonably large (and thus physically unnatural) value of the parameter.
\end{abstract}

\maketitle

\section{Introduction}\label{sec1}

The standard model of cosmology, famously known as the $\Lambda$CDM framework, has provided a robust description of the universe's evolution. Its predictions are highly consistent with extensive high-precision observational data, such as the Cosmic Microwave Background (CMB) anisotropies measured by the Planck satellite~\cite{Planck:2018vyg}, the large-scale structure surveys~\cite{eBOSS:2020yzd}, and the late-time accelerated expansion inferred from Type Ia supernovae observations~\cite{SupernovaSearchTeam:1998fmf,SupernovaCosmologyProject:1998vns}. However, despite its successes, the $\Lambda$CDM model is fundamentally challenged by theoretical puzzles. The most prominent is the ``cosmological constant problem", which highlights the staggering discrepancy between the observed value of vacuum energy and the theoretical predictions from quantum field theory~\cite{Weinberg:1988cp,Carroll:2000fy,Padmanabhan:2002ji}. Furthermore, the recent persistent ``Hubble tension", the $4\sigma$ to $6\sigma$ mismatch between the $H_0$ value measured from the local distance ladder and that inferred from CMB data, suggests that the standard paradigm may require extensions or an overhaul~\cite{Riess:2019cxk,Verde:2019ivm,DiValentino:2021izs}.

In the search for physics beyond $\Lambda$CDM, modified gravity theories have emerged as a promising frontier. Among these, the Randall-Sundrum II (RS-II) braneworld scenario offers a profound geometric perspective by embedding our four-dimensional universe as a 3-brane within a five-dimensional anti-de Sitter bulk~\cite{Randall:1999vf}. A hallmark of the RS-II model is the modification of the Friedmann equation at high energy scales~\cite{Binetruy:1999ut,Binetruy:1999hy,Hebecker:2001nv,Flanagan:1999cu}. These corrections provide a natural mechanism for driving primordial inflation and have been shown to significantly alter the early-universe dynamics, including the thermal history and the production of dark matter~\cite{Maartens:1999hf,Okada:2004nc,Nihei:2004xv,Maartens:2010ar}. Moreover, the nature of the vacuum itself has been scrutinized through the quantum effects in curved spacetime. The running vacuum model (RVM) proposes that the vacuum energy density is not a static constant but a dynamical quantity that evolves as a power series of the Hubble parameter and its time derivatives~\cite{Shapiro:1999zt,Shapiro:2000dz,Shapiro:2003ui,Sola:2005et}. This ``running" is motivated by the renormalization group equations and provides a potential bridge between the high-scale inflation of the early universe and the low-scale dark energy of the current epoch~\cite{Sola:2013gha,Sola:2015rra,SolaPeracaula:2026pgi,SolaPeracaula:2026trz}. The RVM has demonstrated impressive performance in fits to diverse cosmological datasets, often performing better than the rigid cosmological constant term in alleviating the $H_0$ and $\sigma_8$ tensions~\cite{Sola:2016jky,Sola:2017znb,Gomez-Valent:2017idt,Gomez-Valent:2018nib,SolaPeracaula:2018wwm,SolaPeracaula:2021gxi}.

The implementation of inflation in the RS-II model relies on the quadratic energy density term $\rho^2/(2\lambda)$, where $\lambda$ is the brane tension. In the high-energy regime of the early universe, i.e., $\rho \gg \lambda$, the expansion rate $H$ becomes proportional to $\rho$ rather than $\sqrt{\rho}$ as in standard general relativity. This modification results in a significantly enhanced Hubble damping term in the inflaton's equation of motion, $3H\dot{\phi}$. Consequently, the slow-roll conditions are much easier to satisfy, allowing for sufficient inflation even with relatively steep potentials that would fail in the standard four-dimensional cosmological context~\cite{Maartens:1999hf,Hawkins:2000dq,Huey:2001ae,Lidsey:2003ws}. Furthermore, this model predicts distinct signatures in the tensor-to-scalar ratio and the primordial power spectrum, offering a potential observational window into the existence of extra dimensions~\cite{Langlois:2000ns,Zarrouki:2011zoj,Gangopadhyay:2016qqa}. On the other hand, in the RVM, inflation is driven by the higher-order terms in the Hubble parameter $H$, specifically the $H^4$ term, which dominates the early universe. At extremely high energy scales, this term leads to an unstable de Sitter-like phase where $H \approx \text{constant}$, effectively triggering a period of rapid exponential expansion. Unlike standard inflationary models, the RVM does not necessarily require an ad hoc scalar field (inflaton); instead, inflation emerges from the quantum effects of the vacuum itself~\cite{Lima:2013dmf,Sola:2015rra}. A significant advantage of the RVM is its natural mechanism for the ``graceful exit'' and reheating~\cite{Sola:2013gha}. As the universe expands and $H$ decreases, the $H^4$ term loses dominance. As the vacuum energy density decays at the end of the inflationary epoch, it acts as a source term, producing relativistic particles and reheating the universe. 

In this work, by combining the RVM with the RS-II model, we construct a braneworld model in which the brane accommodates running vacuum. We focus on the cosmological evolution of the model, with particular emphasis on the realization of inflation. Moreover, when a dynamic vacuum emerges in cosmological models, a fascinating interaction arises. Because the vacuum energy evolves, the Bianchi identities require a non-vanishing interaction between the vacuum and the other matter, leading to modified conservation equations~\cite{Overduin:1998zv,Shapiro:2009dh,Fritzsch:2012qc,Wang:2016lxa,Salvatelli:2014zta}. This interaction is not merely a dynamical curiosity but is deeply rooted in the thermodynamics of the universe. The second law of thermodynamics and the conditions for thermal equilibrium must be satisfied throughout the cosmic history, especially when particle production occurs due to vacuum decay~\cite{Lima:1995kd,Prigogine:1989zz,Mimoso:2013zhp,Lima:2015mca,SolaPeracaula:2019kfm}. Investigating the entropy evolution in such hybrid models is essential to ensure their physical consistency and to determine whether the dynamical vacuum can lead to a stable thermal state in the late-time limit~\cite{Radicella:2011qpl,Mimoso:2013zhp}. Therefore, we also study the effect of running vacuum on the cosmic entropy evolution in this braneworld model. We consider a special class of solutions and analyze different matter-dominated scenarios, to place constraints on the model. 

The paper is organized as follows: In Sec.~2, we construct a braneworld model that incorporates both the RVM and the RS-II model. We derive a special class of analytical solutions in the context of cosmology. In Sec.~3, we examine the cosmological implications of the model and discuss the constraints on the parameters in the model. Finally, we provide our discussions and concluding remarks in Sec.~4.

\section{Running vacuum on the brane in the RS-II model} 
We begin with the RS-II model with a single 3-brane embedded in an anti-de Sitter bulk. The bulk action is
\begin{equation}
	S = \frac{1}{2\kappa_5^2} \int d^5x \sqrt{-g^{(5)}} \left( R^{(5)} - 2\Lambda_5 \right) + \int d^4x \sqrt{-g} \, \mathcal{L}_{\text{brane}}\,,
\end{equation}
where $\kappa_5^2 = 8\pi G_5$ is the five-dimensional gravitational constant, $\Lambda_5 < 0$ is the bulk cosmological constant, and $\mathcal{L}_{\text{brane}}$ contains the brane tension and the matter fields confined to the brane.

The five-dimensional Einstein equations are given by
\begin{equation}
	G_{AB}^{(5)} + \Lambda_5 \, g_{AB}^{(5)} = \kappa_5^2 \, T_{AB}^{\text{(total)}},
\end{equation}
with $T_{AB}^{\text{(total)}} = T_{AB}^{\text{(bulk)}} + \delta(y) \, T_{\mu\nu}^{\text{(brane)}} \, \delta_A^\mu \delta_B^\nu$, where $y$ is the extra-dimensional coordinate. Using the Gauss-Codazzi equations and the Israel junction conditions, one derives the effective Einstein equation on the brane. For a $\mathbb{Z}_2$-symmetric bulk, the induced four-dimensional Einstein equation takes the form~\cite{Shiromizu:1999wj,Maartens:2010ar}
\begin{equation}
	G_{\mu\nu} = -\Lambda_4 \, g_{\mu\nu} + \kappa_4^2 \, T_{\mu\nu} + \kappa_5^4 \, \Pi_{\mu\nu} - \mathcal{E}_{\mu\nu},
	\label{eq:induced}
\end{equation}
where:
\begin{itemize}
	\item $\Lambda_4 = \frac{\kappa_5^2}{2} \left( \Lambda_5 + \frac{ \lambda^2}{6} \kappa_5^2\right)$ is the effective four-dimensional cosmological constant, which vanishes for the fine-tuned relation $\Lambda_5 = - \lambda^2\kappa_5^4/6$ with $\lambda$ being the brane tension;
	\item $\kappa_4^2 =8\pi G_4= \frac{\lambda }{6} \kappa_5^4$ is the four-dimensional gravitational constant;
	\item $\Pi_{\mu\nu} = \frac{1}{12} T T_{\mu\nu} - \frac{1}{4} T_{\mu\rho} T^\rho_{\;\nu} + \frac{1}{8} g_{\mu\nu} \left( T_{\rho\sigma} T^{\rho\sigma} - \frac{1}{3} T^2 \right)$ is a quadratic correction in the brane matter stress-energy tensor $T_{\mu\nu}$;
	\item $\mathcal{E}_{\mu\nu} = C^{(5)}_{ABCD} \, n^A n^C \, g^B_{\;\mu} g^D_{\;\nu}$ is the projection of the five-dimensional Weyl tensor onto the brane, which encodes the influence of the bulk gravitational field.
\end{itemize}

To study the cosmological implications, we assume a homogeneous and isotropic Friedmann-Lema\^itre-Robertson-Walker (FLRW) metric on the brane:
\begin{equation}
	ds_4^2 = -dt^2 + a(t)^2 \left( \frac{dr^2}{1-kr^2} + r^2 d\Omega^2 \right),
\end{equation}
where $k = -1,0,1$ denotes the spatial curvature. In this work, we only consider flat space ($k=0$). The matter content on the brane is taken as a perfect fluid with energy density $\rho_m$ and pressure $p_m$, so that $T_{\mu\nu} = (\rho_m + p_m) u_\mu u_\nu + p_m g_{\mu\nu}$.

Under these assumptions, the time-time component of the induced equation (\ref{eq:induced}) yields the modified Friedmann equation. The contribution from $\Pi_{\mu\nu}$ gives a term quadratic in $\rho_m$, while $\mathcal{E}_{\mu\nu}$ contributes a ``dark radiation'' term proportional to $a^{-4}$. The result is~\cite{Binetruy:1999ut,Maartens:2000fg,Langlois:2002bb,Brax:2003fv}
\begin{equation}
	H^2  = \frac{\kappa_4^2}{3} \rho_m \left( 1 + \frac{\rho_m}{2\lambda} \right) + \frac{\Lambda_4}{3},
	\label{eq:friedmann}
\end{equation}
where $H = \dot{a}/a$ is the Hubble parameter. Note that the projection of the five-dimensional Weyl tensor onto the brane is assumed to be zero, i.e., the dark radiation term is absent. For the RS fine-tuning $\Lambda_4 = 0$ and in the low-energy limit $\rho_m \ll \lambda$, the standard Friedmann equation is recovered. At high energies $\rho_m \gtrsim \lambda$, the $\rho_m^2$ term dominates, leading to a modified expansion history.

In the standard formulation described above, $\Lambda_{4}$ is strictly a constant. However, insights from quantum field theory in curved spacetime suggest that the vacuum energy should not be static. In the context of the dynamical vacuum, the vacuum energy density $\rho_{\Lambda}(t) \equiv \Lambda_{4}(t)/\kappa_{4}^{2}$ is promoted to a dynamical quantity that evolves with the cosmic expansion. 

Typically, the vacuum energy density is expressed as a series expansion in powers of the Hubble parameter $H$ and its time derivative~\cite{
Shapiro:2000dz,Shapiro:2003ui}:
\begin{equation}
	\rho_{\Lambda}(H) = \frac{\Lambda_4(H)}{\kappa_4^2}= a_{0} + a_{1} H^{2} + a_{2} \dot{H} + \mathcal{O}(H^{4}),
	\label{eq:rvm_ansatz}
\end{equation}
where $a_{i}$ are phenomenological coefficients motivated by the renormalization group running of the vacuum energy. By coupling the dynamical vacuum to the RS braneworld scenario, particularly interacting through the modified Friedmann equation~\eqref{eq:friedmann} and a modified conservation sector, we can explore novel cosmological dynamics from the very early universe to the current dark energy dominated epoch.

In this work, we investigate a widely studied formulation of the dynamical vacuum, which is usually called the running vacuum model (RVM)~\cite{Sola:2013gha,Sola:2015rra,Sola:2016jky,SolaPeracaula:2022hpd}. In the RVM, the dynamical vacuum is given by
\begin{equation}
\rho_{\Lambda}(H) = \frac{\Lambda_4(H)}{\kappa_4^2}= \frac{3}{\kappa_4^2} \left( c_0 + \nu H^2 + \alpha \frac{H^4}{H_I^2} \right).\label{rvmformulation}
\end{equation}
Here, $c_0$ is a constant with a dimension of energy squared, which is used to recover the $\Lambda$CDM model with $\nu=0=\alpha$. The dimensionless coefficients $\nu$ and $\alpha$ can be constrained by cosmological observations. 

Taking Eq.~\eqref{rvmformulation} into Eq.~\eqref{eq:friedmann} yields the modified Friedmann equation in the RS-II model with running vacuum:
\begin{equation}
H^2 = \frac{\kappa_4^2}{3} \rho_m \left( 1 + \frac{\rho_m}{2\lambda} \right) + c_0 + \nu H^2 + \alpha\frac{H^4}{H_I^2}.\label{modfriedm}
\end{equation}
On the other hand, the contracted Bianchi identity, $\Delta_{\mu}G^{\mu\nu}=0$, requires the matter energy-momentum tensor $T_m^{\mu\nu}$ to satisfy
\begin{equation}
\nabla_\mu T_m^{\mu\nu} = g^{\mu\nu} \nabla_\mu \rho_\Lambda(H).
\end{equation}
Using the FLRW metric and the definition of the matter energy-momentum tensor, it can be rewritten as  
\begin{equation}
\dot{\rho}_m + 3H(\rho_m + p_m) = -\frac{\dot{\Lambda}_4(H)}{\kappa_4^2}=-\frac{3}{\kappa_4^2}\left(2\nu H\dot H+4\frac{\alpha}{H_I^2} H^3\dot H  \right).\label{modconver}
\end{equation}
Once the matter is characterized by a specific equation of state (EoS), the time evolution of the scale factor can be fully determined by Eqs.~\eqref{modfriedm} and~\eqref{modconver}. 

In this work, we explore the viability of inflation in a scenario where the quadratic density term ($\rho_m^2$) from the RS-II model and the quartic Hubble term ($H^4$) from the RVM mutually cancel. The motivation for this configuration is twofold. First, it facilitates the derivation of exact analytical solutions, thereby streamlining the parameter space and enabling more efficient constraints. Second, by omitting the terms responsible for inflation from both models and adopting modified solutions, we aim to investigate the model's capacity to provide a unified evolution of the universe, spanning from the early inflationary epoch to the present era under such circumstance. In this case, we first present a special class of analytical solutions for the RS-II model with running vacuum. To simplify the problem, we assume that the universe, in addition to running vacuum, consists of only a single perfect fluid with an EoS $p=\omega\rho$, where $\omega$ is a constant. It is found that if the $H^4$ term in running vacuum is strictly equal to the square of the matter energy density ($\rho_m^2$) and $c_0=0$ (where $c_0=0$ implies the absence of a constant term in the vacuum energy, which is then entirely governed by the Hubble parameter), a set of simplified analytical solutions can be obtained. This requirement effectively splits Eq.~\eqref{modfriedm} into the following two equations:
\begin{align}
&(1+\nu)H^2=\frac{\kappa_4^2}{3} \rho_m, \label{eq11}
\\
&\frac{\kappa_4^2}{3} \frac{\rho_m^2}{2\lambda}=\alpha\frac{H^4}{H_I^2}.
\end{align}
Note that the deviation of Eq.~(\ref{eq11}) from the standard Friedmann equation seems to be parameterized only by $\nu$, but the solutions to the equation are also influenced by the imposed constraints. With these two equations, we can obtain
\begin{equation}
\frac{\alpha}{H_I^2}=\frac{3(1+\nu)^2}{2\lambda\,\kappa_4^2}.\label{eq13}
\end{equation}
This equation implies that the parameters in the model are no longer entirely independent. By utilizing this relation, we can replace all terms involving $\frac{\alpha}{H_I^2}$ with $\frac{3(1+\nu)^2}{2\lambda\,\kappa_4^2}$. Substituting Eq.~\eqref{eq11} into Eq.~\eqref{modconver}, we obtain
\begin{equation}
\frac{18 (1+\nu)^2 }{\lambda\,\kappa_4^2}H^2 \dot H+(6+12 \nu) \dot H+9 (1+\nu) (1+\omega) H^2=0.
\end{equation}
For an arbitrary value of $\omega$, the solution to the aforementioned equations is given by
\begin{equation}
H(t)=\frac{ \sqrt{3{\tilde\lambda} (1+\nu)^2 \left[3 {\tilde\lambda}  (1+\omega)^2\,t^2+32 v+16\right]}-3 {\tilde\lambda} (1+\nu) (1+\omega)\,t}{12 (1+\nu)^2},
\end{equation}
where ${\tilde\lambda}=\lambda\,\kappa_4^2$ for brevity. With this analytical solution, we can explicitly investigate whether universes dominated by different forms of matter satisfy the fundamental observational requirements of cosmic evolution. Furthermore, it enables us to constrain the parameters in the model.

\section{Constraints on the RS-II model with running vacuum from different matter-dominated universes}

In this section, we investigate how to utilize observations of cosmic evolution to constrain both the RVM and the RS-II model under various matter-dominated scenarios. As previously established using the constraint in Eq.~\eqref{eq13}, the coefficient $\alpha$ (associated with the $H^4$ term) in the RVM has been replaced by the coefficient $\nu$ (associated with the $H^2$ term). Consequently, in our subsequent analysis, constraining the RVM is equivalent to constraining the parameter $\nu$. Meanwhile, constraining the RS-II model refers to placing bounds on the brane tension and the bulk cosmological constant. We primarily examine whether each model satisfies the expansion history of the universen—specifically, the existence of a primordial inflationary epoch, a decelerated expansion phase, and a late-time acceleration period. By imposing these requirements, we can derive straightforward constraints on the model parameters. This analytical solution provides a unified framework to address the key stages of cosmic evolution, including primordial inflation, the graceful exit mechanism, and the late-time accelerated expansion. Furthermore, the evolution of cosmic entropy under this solution is shown to be consistent with fundamental thermodynamic requirements. By successfully connecting these distinct epochs, the model offers a self-consistent description of the universe's entire dynamical history.

\subsection{Dust-dominated universe}
In the standard $\Lambda$CDM framework, the energy density of the present-day universe is primarily composed of non-relativistic matter (dust and dark matter) and the cosmological constant (vacuum energy). When both components are incorporated into the Friedmann equation, the model accurately describes the late-time evolution of the universe—a stage where vacuum energy gradually supersedes matter as the dominant factor governing the expansion rate, driving the universe into an accelerated expansion phase. Therefore, for a dust-dominated cosmological model, it is strictly only required to account for this late-time acceleration. 

In this work, to demonstrate the advantages of our proposed model, we consider its complete evolutionary history and investigate the conditions under which it reproduces the dynamical expansion trajectory of the actual universe, thereby deriving constraints on the relevant parameters. Furthermore, due to the evolution of the vacuum, the energy-momentum tensor of the matter field—which is non-minimally coupled to running vacuum—is generally not conserved, leading to modifications in the thermodynamic properties of matter. We take into account the influence of running vacuum on matter entropy and require the system to satisfy the second law of thermodynamics to further constrain the associated parameters.

For ideal dust, the EoS is $\omega=0$. Consequently, the Hubble parameter of the model is given by
\begin{equation}
H(t)=\frac{ \sqrt{3{\tilde\lambda} (1+\nu)^2 \left(3 {\tilde\lambda} \,t^2+32 v+16\right)}-3 {\tilde\lambda} (1+\nu) t}{12 (1+\nu)^2}.\label{4.11}
\end{equation}
First, we must ensure that the universe undergoes continuous expansion, which imposes the condition $H'(t) > 0$. Given the relation $\kappa_4^2 = 8\pi G_4 = \frac{\lambda}{6} \kappa_5^4$, it follows that $\tilde{\lambda} > 0$. Consequently, a preliminary constraint on the parameter $\nu$ can be established as $\nu > -1/2$.

From a dynamical standpoint, the fundamental requirement for the model to achieve inflation is that the first slow-roll parameter, $E(t) = -\frac{H'(t)}{H(t)^2}$, must satisfy $E(t) < 1$. Within the time interval $[t_1, t_2]$ during which $E(t) < 1$ holds, we evaluate whether the number of e-folds, defined as $N_k = \int_{t_1}^{t_2} H(t) dt$, exceeds the standard threshold of 50–60. The inflationary phase terminates at $t_2$ when the condition $E(t) = 1$ is met. Subsequently, the universe enters a prolonged period of decelerated expansion until the onset of late-time acceleration at $t_3$. In the following, we examine whether the Hubble parameter satisfies these evolutionary criteria and, consequently, derive further constraints on the parameters $\tilde\lambda$ and $\nu$.

According to Eq.~\eqref{4.11}, the condition that the first slow-roll parameter $E(t) = -\frac{H'(t)}{H(t)^2} < 1$ is equivalent to
\begin{equation}
E(t)=\frac{3}{4\,(1+2\nu)}\left( 1 + \nu + \dfrac{\sqrt{3}\,\tilde\lambda\,(1+\nu)^2\,t}{\sqrt{\tilde\lambda\,(1+\nu)^2\left(16 + 3\,\tilde\lambda\,t^2 + 32\nu\right)}} \right)<1.
\label{4.12}
\end{equation}
It can be demonstrated that in the limit $t \to 0$, the condition for accelerated expansion, $E(t) < 1$, is satisfied only if $\nu > -1/5$. Furthermore, for a fixed early time $t$, $E(t)$ is a monotonically decreasing function of $\nu$. However, it is crucial to note that even when $\nu > -1/5$, it is impossible to achieve the standard slow-roll condition, $E(t) \ll 1$. This is because in the limit $\nu \to \infty$, the parameter $E(t\to 0)$ approaches a minimum value of $3/8$. Consequently, this model cannot realize conventional slow-roll inflation.

We therefore investigate the possibility of non-slow-roll inflation, which requires the calculation of the number of e-folds. For such a scenario to be viable, the condition $\nu > -1/5$ must still be imposed. This ensures an initial phase of acceleration followed by a period of deceleration, thereby providing a natural mechanism for a graceful exit from inflation.

\begin{figure}[!htb]
\center{
\includegraphics[width=11cm]{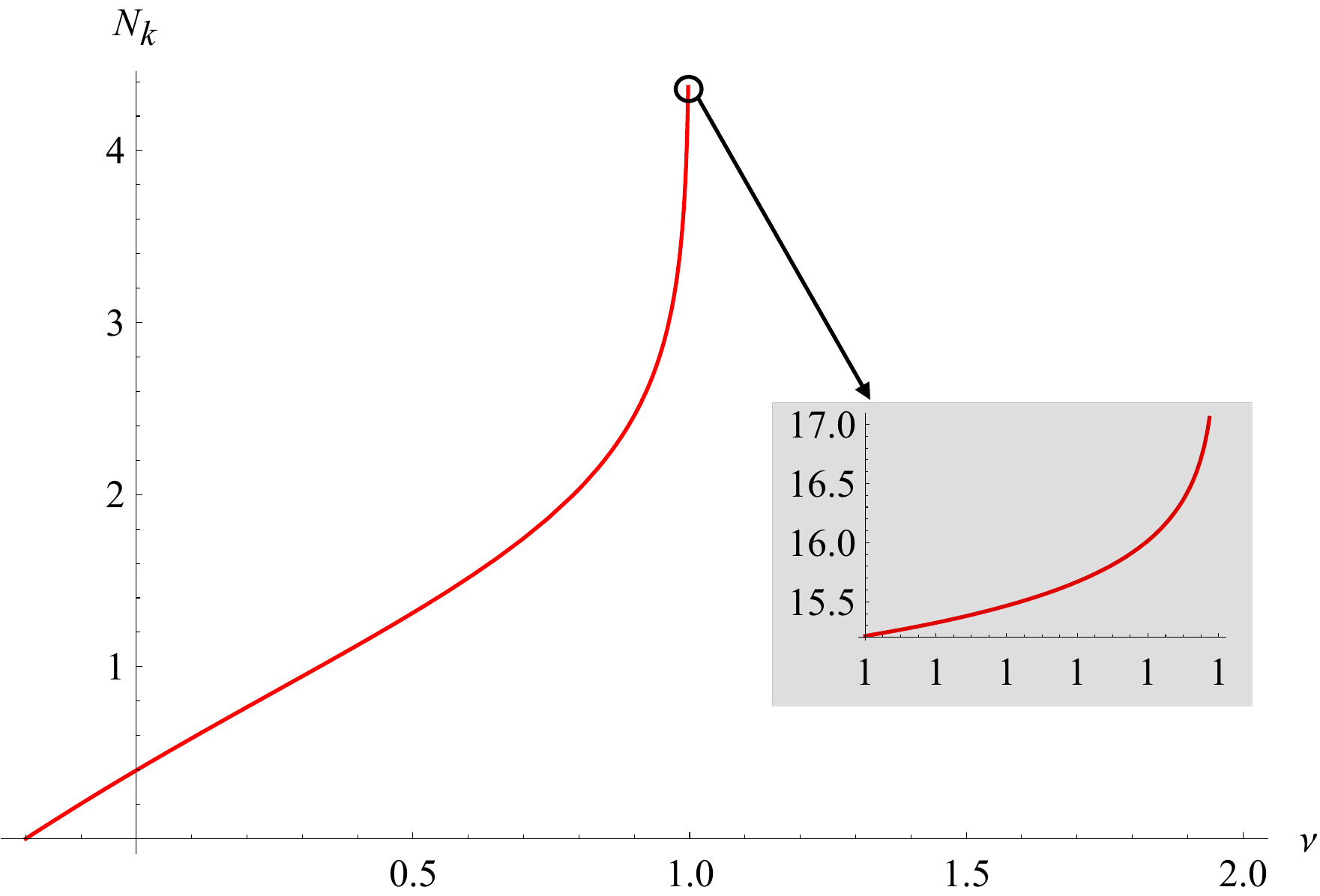}
}
\caption{Variation of the e-folds number $N_{k}$ with the parameter $\nu$ in dust-dominated universe.}
\label{fig1}
\end{figure}

First, we identify the point where $E(t) = 1$, which marks the onset of decelerated expansion; it is given by
\begin{equation}
t_2= \sqrt{\frac{2(1+5 \nu)^2}{3\tilde\lambda (1-\nu)}}.
\end{equation}
As observed here, the parameter $\nu$ must be strictly less than unity ($\nu < 1$); otherwise, a physically meaningful value for $t_2$ does not exist. Starting from the initial time $t_1 = 0$, the number of e-folds is calculated as
\begin{equation}
N_{k}=\frac{\nu (5+\log 16)+4(1+2\nu) /\text{arctanh}\left(\frac{3\nu}{\nu+2}\right)+1+\log4}{6 (\nu+1)}.
\end{equation}
As shown here, the e-folds number $N_{k}$ is independent of the brane tension $\tilde\lambda$. We illustrate the relationship between the e-folds number $N_{k}$ and the parameter $\nu$ in Fig.~\ref{fig1}. It is evident that for $\nu$ within the interval $(-1/5, 1)$, it is difficult for the e-folds number $N_{k}$ to exceed the required threshold of 50--60. Based on the properties of the $\text{arctanh}$ function, although $N_{k}$ can theoretically diverge as $\nu \to 1$ (see the enlarged portion in Fig.~\ref{fig1}, where the range of the horizontal axis is actually from $1-10^{-12}$ to 1), achieving a value of $N_{k} \in [50, 60]$ would require $\nu$ to be fine-tuned to dozens of decimal places (such as $N_{k}\sim 59$ for $\nu=1-{10^{-50}}$). Such an extreme degree of fine-tuning is physically undesirable and difficult to justify.

Furthermore, it can be demonstrated that for $t > 0$ and $\nu > 1$, the universe undergoes eternal accelerated expansion. Consequently, a contradiction arises in this model between the requirement for late-time acceleration and the necessity of a graceful exit from inflation. It can be concluded that this dust-dominated universe does not accommodate a realistic inflationary process. Instead, a universe dominated by matter and running vacuum is better suited for describing the decelerated expansion phase following inflation, which is essential for large-scale structure formation, provided that the parameter $\nu$ satisfies $-1/5 < \nu < 1$.

We now examine the evolution of dust entropy during this epoch. Following the framework established in Ref.~\cite{SolaPeracaula:2019kfm}, we neglect any variations in the rest mass of the dust particles and focus exclusively on the process of particle production. By utilizing the aforementioned analytical solutions and the EoS of dust, the particle production rate is expressed as
\begin{equation}
\begin{split}
\Gamma(t)=&\frac{\dot N}{N}=\frac{\dot{\rho}_d + 3H(\rho_d + p_d)}{(\rho_d + p_d)}=-\frac{ \dot{\Lambda}_4(H)}{3 (1+\nu) H^2}\\
=& \frac{3\sqrt{3}\, \tilde\lambda^2\, t^2-3\tilde\lambda f(t)\, t+24\sqrt{3}\, \tilde\lambda \,\nu+8\sqrt{3}\, \tilde\lambda}{4 (1+\nu)f(t)}.
\end{split}
\end{equation}
where $f(t)=\sqrt{3{\tilde\lambda}^2\, t^2+32 {\tilde\lambda}\,\nu+16{\tilde\lambda}}$\,.

Assuming that the specific entropy $\sigma_0$ (the entropy per particle) remains constant, the total entropy of the dust particles within a comoving volume can be expressed as
\begin{equation}
\begin{split}
\frac{\dot {S}(t)}{S(t)}=\frac{\dot N \sigma_0}{N\sigma_0}=\Gamma(t).
\end{split}
\end{equation}
Assuming that the entropy of particles within a comoving volume at the initial time is $S_0$, the entropy at any arbitrary time is given by
\begin{equation}
\begin{split}
S(t)=&S_0\exp\left[\int_{t_0}^{t}\Gamma(t)\, dt\right]\\
=&S_0\exp \left[\frac{16\, \nu \log \left(\sqrt{3} f(t)+3 \tilde\lambda \,t\right)+\sqrt{3} f(t)\, t-3 \tilde\lambda\, t^2}{8 (1+\nu)}-C_0\right].
\end{split}
\end{equation}
Here, $c$ is the corresponding integration constant. The second law of thermodynamics requires $\dot{S}(t) > 0$, which is equivalent to the condition $\Gamma(t) > 0$. It can be demonstrated that for $\dot{S}(t) > 0$ to hold for all $t$, the parameter $\nu$ must satisfy $\nu > 0$. 

Next, we examine whether the entropy in this dust-dominated cosmological model eventually approaches a final stable state. This investigation simply amounts to determining whether $S(t)$ diverges as $t \to \infty$. Based on the evolution of the entropy, it can be calculated that $S(t \to \infty)$ is infinite; therefore, the dust entropy in this model does not reach a state of thermal equilibrium. However, considering that this model is best suited for describing the decelerated expansion phase of the universe, it is sufficient to satisfy the second law of thermodynamics. During this period, the constraint on the parameter $\nu$ in the RVM can be summarized as $0 < \nu < 1$. While the model can theoretically achieve inflation, it would necessitate a significant degree of fine-tuning.

\subsection{Radiation-dominated universe}

For a cosmological model consisting of radiation and running vacuum, we continue to examine the complete evolutionary history of the universe to verify whether it fulfills the aforementioned requirements for the dynamical expansion trajectory. In this scenario, the analytical solution for the Hubble parameter is given by
\begin{equation}
H(t)=\frac{ \sqrt{{\tilde\lambda} (1+\nu)^2 \left({\tilde\lambda} \,t^2+6 \nu+3\right)}-{\tilde\lambda} (1+\nu) t}{3 (1+\nu)^2}.\label{4.21}
\end{equation}
Under this solution, to ensure that the universe remains in a state of continuous expansion, and in conjunction with the condition $\tilde{\lambda} > 0$, the preliminary constraint on the parameter $\nu$ remains $\nu > -1/2$.

In this model, the first slow-roll parameter $E(t) = -\frac{H'(t)}{H(t)^2}$ is given by
\begin{equation}
E(t)=\frac{3 (1+\nu)}{\tilde\lambda\, t^2-t \sqrt{\tilde\lambda \left(\tilde\lambda\, t^2+6 \nu+3\right)}+6 \nu^2+9 \nu+3}.
\label{4.12}
\end{equation}
It is evident that as $t \to 0$, the condition $E(t) < 1$—which is necessary for the existence of accelerated expansion—is satisfied only when $\nu > 0$. Similar to the previous case, the value of $E(t)$ decreases as $\nu$ increases. It should be noted that for $\nu > 0$, the deep slow-roll regime $E(t) \ll 1$ cannot be attained, as even in the limit $\nu \to \infty$, the minimum value of $E(t \to 0)$ is $1/2$. Consequently, this model is also incapable of realizing standard slow-roll inflation. We now investigate whether a non-slow-roll inflationary phase can occur.

\begin{figure}[!htb]
\center{
\includegraphics[width=11cm]{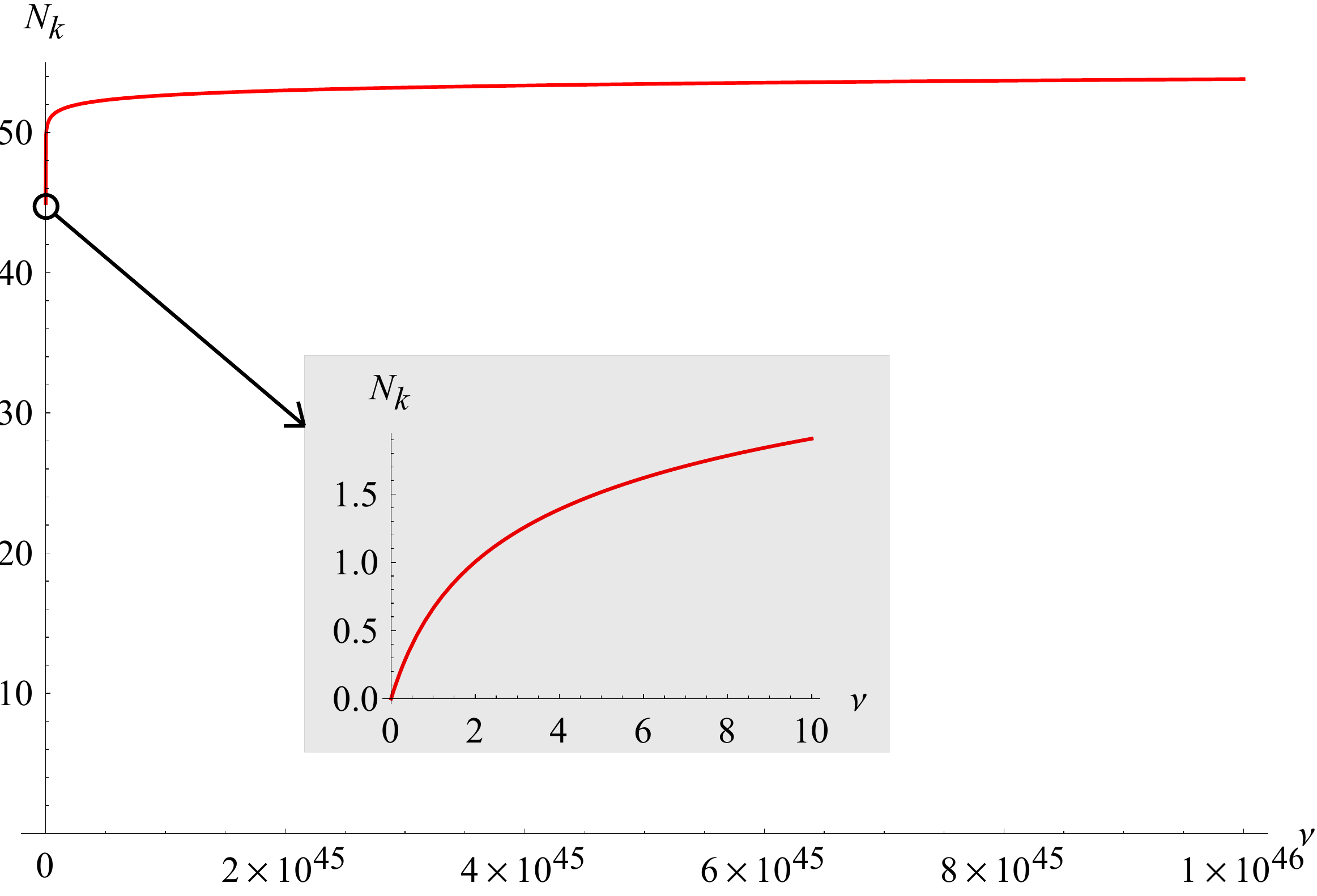}
}
\caption{Variation of the e-folds number $N_{k}$ with the parameter $\nu$ in radiation-dominated universe.}
\label{fig2}
\end{figure}

We first identify the point where $E(t)=1$. The result of the calculation is given by
\begin{equation}
t_2= \sqrt{\frac{{3} }{\tilde\lambda}}\,\nu.
\end{equation}
As observed here, unlike the dust-dominated model, there are no additional constraints on the parameter $\nu$ in this case. Starting from the initial time $t_1 = 0$, the number of e-folds is given by
\begin{equation}
N_{k}=\frac{2 \nu+(1+2\nu) \log (1+2 \nu)}{4 (1+\nu)}.
\end{equation}
It is evident that the number of e-folds remains independent of the brane tension. We present the relationship between the e-folds number $N_{k}$ and the parameter $\nu$ in Fig.~\ref{fig2}. It is noted that when $\nu$ is not sufficiently large, it is difficult for the e-folds number $N_{k}$ to exceed the required threshold of 50--60. However, an analysis of the functional properties reveals that $N_{k}$ can diverge as $\nu \to \infty$. We find that for $\nu$ on the order of $10^{45}$, $N_{k}$ can reach 50. Unlike the previous scenario which required extreme fine-tuning, this model necessitates an enormous value for $\nu$, which is likewise considered physically unnatural.

Furthermore, it can be demonstrated that for $t > 0$ and $\nu > 0$, no value of $\nu$ exists that can ensure the universe remains in a state of eternal accelerated expansion. While $\nu > 0$ guarantees an initial period of acceleration, the universe will inevitably transition into a decelerated expansion phase. Therefore, it is generally concluded that this radiation-dominated universe does not possess a realistic inflationary process. Similar to the dust-dominated case, a universe governed by radiation and running vacuum is better suited for describing the decelerated expansion phase near the end of inflation and the subsequent era of structure formation, provided that the parameter $\nu$ satisfies $\nu > 0$.

During this epoch, given the analytical solutions for the scale factor and the Hubble parameter, and according to the definition of radiation entropy within a comoving volume, we obtain
\begin{equation}
\begin{split}
S(t)=&\frac{p_r+\rho_r}{T_r}a^3=\frac{4}{3}\left(\frac{\pi^2}{30}g_*\right)^{1/4} \left(\frac{18 (1+\nu)}{\lambda\, \kappa_5^4}\right)^{3/4} a(t)^3 H(t)^{3/2}\\
=&C_0\left(B(t)-\tilde\lambda\, t\right)^{3/2}  \left(B(t)+\tilde\lambda\, t\right)^{3-\frac{3}{2 (1+\nu)}}\exp\left[{\frac{B(t)\,t-\tilde\lambda\, t^2}{2 (1+\nu)}}\right],
\end{split}
\end{equation}
where we have used $\rho_r(t)=\frac{18(1+\nu)H(t)}{\lambda\, \kappa_5^4}=\frac{\pi^2}{30}g_*T_r^4$ to simplify the expression. Here, $B(t)=\sqrt{ \tilde\lambda^2 t^2+6 \tilde\lambda\,\nu+3\tilde\lambda}$ and $C_0$ is a positive constant depending on the parameters $\nu$, $g_*$, $\tilde\lambda$, and $\kappa_5^4$. Moreover, $g_*$ represents the total number of effectively massless degrees of freedom~\cite{Kolb:1990vq}, and $T_r$ is the radiation temperature. The second law of thermodynamics requires $\dot{S}(t) > 0$, which, according to the expression above, is equivalent to
\begin{equation}
\begin{split}
2\nu\, \tilde\lambda\, t^2 +2\nu\,B(t)\, t +18\nu^2+15\nu+3>0.
\end{split}
\end{equation}
Since $\nu > 0$, it follows that the entropy is strictly monotonically increasing. Given that the entropy diverges as $t \to \infty$, the radiation entropy within a comoving volume does not approach a state of thermal equilibrium. However, considering that this model is primarily intended to describe the decelerated expansion phase of the universe, fulfilling the second law of thermodynamics is sufficient. Consequently, the constraint on the RVM coupling parameter for this period can be summarized as $\nu > 0$. Although this model can theoretically realize an inflationary phase, it would necessitate an exceptionally large value for the parameter $\nu$.

\subsection{General perfect fluid-dominated universe}
Based on the results obtained above, the evolution of a universe composed of running vacuum and various matter species is significantly influenced by the cosmic equation of state, which in turn imposes distinct requirements on the vacuum parameter $\nu$. We have observed that for both dust-dominated and radiation-dominated scenarios, it is impossible to naturally satisfy the complete dynamical expansion trajectory of the universe; these models can only accommodate the inflationary and decelerated expansion phases under specific parameter fine-tuning. Furthermore, although late-time acceleration is achievable in both cases, the corresponding allowed regions for the parameter $\nu$ do not overlap. On the thermodynamic front, while the second law of thermodynamics is generally satisfied, the entropy fails to converge toward an equilibrium state. 

We now consider a universe containing a general perfect fluid characterized by the EoS $P=\omega\rho$. Since the correspondence between a general perfect fluid and arbitrary cosmological phases is not well-defined, nor is the calculation of its entropy, we restrict our focus to whether it can realize inflationary expansion. In this general case, the analytical solution for the Hubble parameter is given by:
\begin{equation}
H(t)=\frac{ (1+\nu)\sqrt{3{\tilde\lambda} \left[3(1+\omega)^2{\tilde\lambda} \,t^2+32 \nu+16\right]}-3{\tilde\lambda}(1+\omega) (1+\nu)t}{12 (1+\nu)^2}.\label{4.21}
\end{equation}
The requirement for cosmic expansion is $\dot{H} > 0$. For $t > 0$, this condition is satisfied provided that $3 \lambda \kappa_4^2 (1 + \nu)^2 (16 + 32 \nu) > 0$. Given that the brane tension $\tilde\lambda$ is strictly positive, this reduces to $16 + 32 \nu > 0$, which implies $\nu > -1/2$. This result confirms that the requirement for expansion is indeed independent of the EoS parameter $\omega$.

The condition for accelerated expansion is $\dot{H} + H^2 > 0$. For the model to accommodate an early-time inflationary epoch, the universe must undergo accelerated expansion in the limit $t \to 0$. From this, we find that although the condition $\omega < -1$ would always ensure initial acceleration, we exclude such a phantom-like scenario from our consideration. Instead, we restrict our analysis to the physically motivated range $\omega\geq -1$. Under this assumption, we have 
\begin{equation}
-1\leq \omega<\frac{5 \nu+1}{3 \nu+3}.
\end{equation}
It is evident that the lower bound for $\omega$ is $-1$ (at $\nu = -1/2$), while the upper bound is $5/3$ (in the limit $\nu \to \infty$). This implies that for $\nu > -1/2$, $\omega$ in the range $(-1, 5/3]$ can guarantee the initial accelerated expansion of the universe, provided that $\nu$ and $\omega$ satisfy the aforementioned relation. This suggests that by tuning the equation-of-state parameter $\omega$, it might be possible to achieve inflation, or even the slow-roll inflationary regime. 

For the late-time universe, as $t \to \infty$, the condition required to ensure accelerated expansion is $-1 < \omega < \frac{\nu-1}{3(1+\nu)}$. Since the upper bound of this expression is $1/3$, it follows that radiation ($\omega=1/3$) cannot sustain eternal acceleration regardless of the value of $\nu$. Meeting this condition ensures that the universe remains in a state of eternal accelerated expansion. To investigate whether inflation can be realized, we now assume $-1 < \omega < \frac{1+5\nu}{3(1+\nu)}$ to guarantee the existence of an initial acceleration phase.

\begin{figure}[!htb]
\center{
\includegraphics[width=11cm]{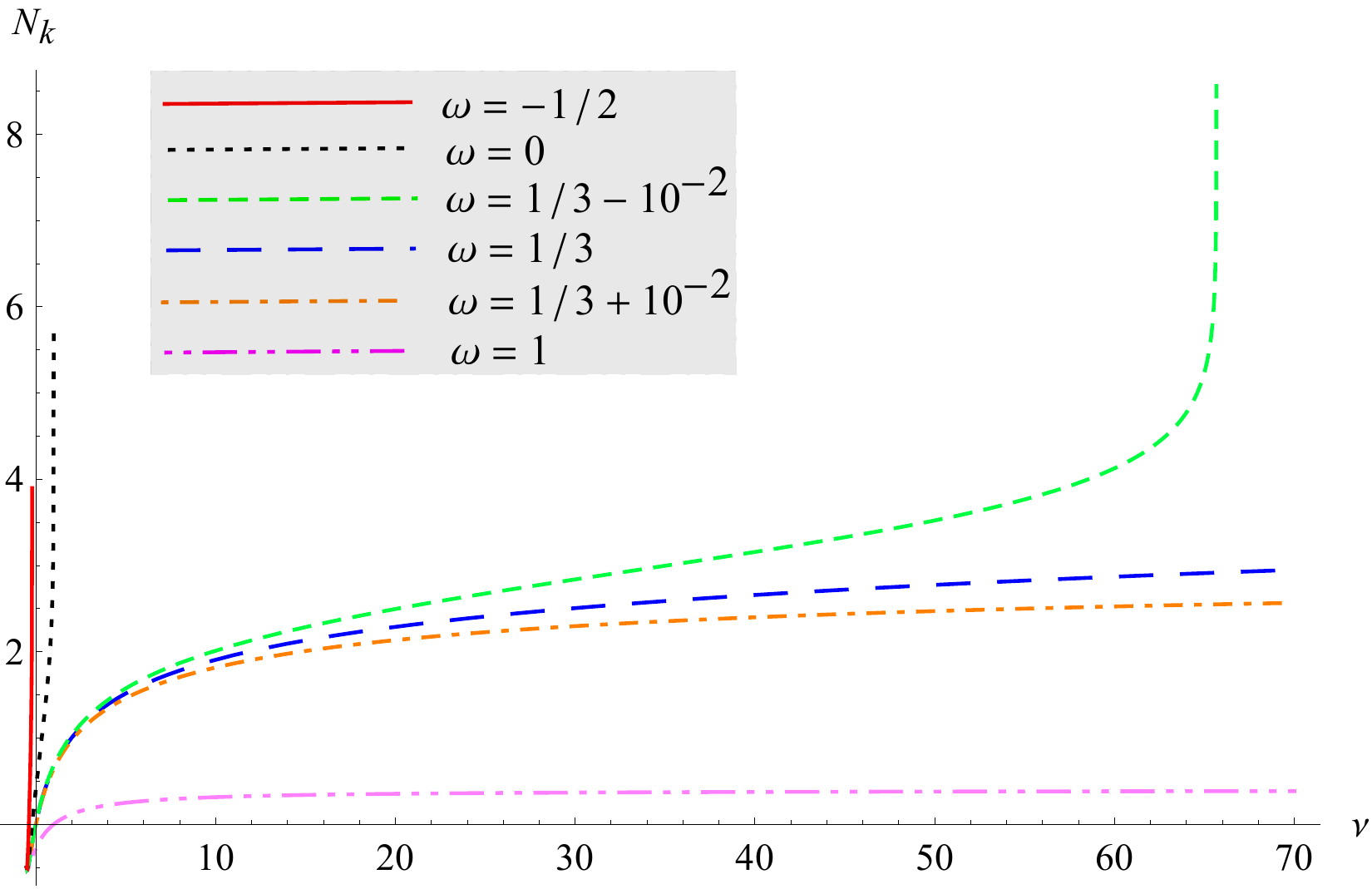}
}
\caption{Variation of the e-folds number $N_{k}$ with the parameter $\nu$ in general perfect fluid-dominated universe.}
\label{fig3}
\end{figure}

For $\nu>-1/2$, the condition $\omega < \frac{1+5\nu}{3(1+\nu)}$ is less restrictive than $\omega < \frac{\nu-1}{3(1+\nu)}$. We now proceed to investigate whether an inflationary phase, followed by a graceful exit, can be realized within the parameter space defined by $-1 < \omega < \frac{1+5\nu}{3(1+\nu)}$. To begin, we examine the first slow-roll parameter:
\begin{equation}
E(t)=\frac{12 (1+\nu) (1+\omega)}{(1+\omega) \left(3 \tilde\lambda  (1+\omega)\,t- \sqrt{3{\tilde\lambda} \left[3(1+\omega)^2{\tilde\lambda} \,t^2+32 \nu+16\right]}\right)t+32\nu+16}.
\end{equation}
It follows that the onset of the decelerated expansion phase occurs at the time $t$, which is given by
\begin{equation}
t_2= \frac{\sqrt{2} \left[1+5 \nu-3 (1+\nu) \omega\right]}{\sqrt{3 \tilde\lambda} \,(1+\omega) \sqrt{1-\nu+3 (1+\nu) \omega}}.
\end{equation}
In this context, we impose a further requirement on $\omega$ and $\nu$ such that $1-\nu+3 (1+\nu) \omega > 0$. Combining this with the aforementioned conditions, we obtain the following constraint on the EoS parameter:
\begin{equation}
    \frac{\nu-1}{3(1+\nu)} < \omega < \frac{1+5\nu}{3(1+\nu)}.
\end{equation}
As $\nu \to -1/2$, the lower bound of this interval remains $-1$, while in the limit $\nu \to \infty$, the upper bound reaches $5/3$. Under these conditions, the number of e-folds, calculated from the initial time $t_1 = 0$, is similarly given by:
\begin{equation}
\begin{split}
N_{k}=&\frac{1}{6 (1+\nu) (1+\omega)}\Big[1+\log (4)+5\nu-3 \omega-3 \nu\, \omega+2 \log (2 \nu+1)\\
&+4 \nu \log (2+4\nu)-2 (1+2\nu) \log (1+3\omega+3\nu\,\omega-\nu)\Big].
\end{split}
\end{equation}
It is also observed that the e-folds number $N_k$ remains independent of $\tilde{\lambda}$. We plotted the dependence of $N_k$ on the parameter $\nu$ for various values of $\omega$ in Fig.~\ref{fig3}. 

It is evident that achieving a sufficient inflationary phase remains a formidable challenge, even when these parameters are adjusted over a wide range. It is found that $\omega = 1/3$ acts as a critical threshold. Specifically, if $\omega < 1/3$, an extreme degree of fine-tuning is required—extending to dozens of decimal places for $\nu$ to yield a value of $N_{k} > 50$. Conversely, if $\omega \geq 1/3$, $\nu$ must take an exceptionally large and physically unnatural value to satisfy the same $N_{k} > 50$ requirement. Such conditions are generally considered physically unrealistic. Consequently, regardless of the matter content, it is difficult for this model to accommodate a natural inflationary process.

\section{Conclusions and discussion} 

In this paper, we have explored the cosmological implications of the RS-II braneworld model where the brane accommodates running vacuum. By assuming a specific configuration where the quadratic energy density term ($\rho_m^2$) from the RS-II model and the quartic Hubble term ($H^4$) from the RVM mutually cancel, we derived a class of exact analytical solutions. This framework allows for a unified description of the universe's evolution from the early inflationary epoch to the late-time acceleration. 

In both dust-dominated and radiation-dominated scenarios, the model struggles to naturally realize a realistic inflationary phase. For the dust case, the parameter $\nu$ must be fine-tuned to an extreme degree (dozens of decimal places) to achieve $N_k \sim 60$. For the radiation case, $\nu$ must take an enormous value (of order $10^{45}$), which lacks a clear physical justification. In general perfect fluid-dominated universes, it is found that the feasibility of inflation is highly sensitive to the EoS parameter $\omega$. We identified $\omega = 1/3$ as a critical threshold. For $\omega < 1/3$, the model is hindered by the fine-tuning problem, while for $\omega \ge 1/3$, the required parameter space for $\nu$ is physically unnatural. Despite the difficulties in achieving inflation, the model is basically thermodynamically consistent in  dust-dominated and radiation-dominated universe. For dust, the condition $\dot S>0$ translates into $\nu>0$, which overlaps with the allowed range for decelerated expansion. For radiation, $\nu>0$ automatically guarantees positive entropy production. However, in both cases the comoving entropy diverges as $t\to\infty$, indicating that the universe never settles into a thermal equilibrium state. This suggests that the particle production rate remains active indefinitely, which is physically implausible in the very late universe; a more complete model should include a mechanism that shuts off the vacuum decay at late times. 

Finally, the present model fails to incorporate the effects of the brane tension  $\lambda$ within both the thermodynamic and inflationary regimes. Future work could use the analytical background solutions to compute the power spectra and compare with Planck or DESI data, thereby placing robust constraints on $\nu$ and the brane tension $\lambda$. In particular, the radiation-dominated branch with $\nu>0$ affects the early expansion rate and could leave signatures in the damping tail of the CMB or in the abundance of light elements.

\vspace{10pt}

\noindent {\bf Acknowledgments}

\noindent This work was supported by the National Natural Science Foundation of China (Grants No. 12505057 and No. 12505058), and the China Postdoctoral Science Foundation (Grant No. 2024M753825).

\end{document}